\documentclass[longauth]{aa}
\usepackage{graphicx}
\usepackage{natbib}
\bibpunct{(}{)}{;}{a}{}{,} 

\begin{document}

\newcommand{\Cerenkov}{Cherenkov\ }
\newcommand{\HMS}[3]{$#1^{\mathrm{h}}#2^{\mathrm{m}}#3^{\mathrm{s}}$}
\newcommand{\DMS}[3]{$#1^\circ #2' #3''$}
\newcommand{\TODO}[1]{\textbf{TODO: \emph{#1}}}
\renewcommand{\cite}{\citep} 

\title{Discovery of a point-like very-high-energy $\gamma$-ray source in Monoceros} 

\titlerunning{A point-like $\gamma$-ray source in Monoceros} 
\authorrunning{F. A. Aharonian et al.}

\author{F. A. Aharonian\inst{1,13}
 \and A.G.~Akhperjanian \inst{2}
 \and A.R.~Bazer-Bachi \inst{3}
 \and B.~Behera \inst{14}
 \and M.~Beilicke \inst{4}
 \and W.~Benbow \inst{1}
 \and D.~Berge \inst{1} \thanks{now at CERN, Geneva, Switzerland}
 \and K.~Bernl\"ohr \inst{1,5}
 \and C.~Boisson \inst{6}
 \and O.~Bolz \inst{1}
% ...
}

\institute{
Max-Planck-Institut f\"ur Kernphysik, P.O. Box 103980, D 69029
Heidelberg, Germany
\and
 Yerevan Physics Institute, 2 Alikhanian Brothers St., 375036 Yerevan,
Armenia
\and
Centre d'Etude Spatiale des Rayonnements, CNRS/UPS, 9 av. du Colonel Roche, BP
4346, F-31029 Toulouse Cedex 4, France
\and
Universit\"at Hamburg, Institut f\"ur Experimentalphysik, Luruper Chaussee
149, D 22761 Hamburg, Germany
\and
Institut f\"ur Physik, Humboldt-Universit\"at zu Berlin, Newtonstr. 15,
D 12489 Berlin, Germany
\and
LUTH, UMR 8102 du CNRS, Observatoire de Paris, Section de Meudon, F-92195 Meudon Cedex,
France
\and
DAPNIA/DSM/CEA, CE Saclay, F-91191
Gif-sur-Yvette, Cedex, France
\and
University of Durham, Department of Physics, South Road, Durham DH1 3LE,
U.K.
\and
Unit for Space Physics, North-West University, Potchefstroom 2520,
    South Africa
\and
Laboratoire Leprince-Ringuet, IN2P3/CNRS,
Ecole Polytechnique, F-91128 Palaiseau, France
\and 
Laboratoire d'Annecy-le-Vieux de Physique des Particules, IN2P3/CNRS,
9 Chemin de Bellevue - BP 110 F-74941 Annecy-le-Vieux Cedex, France
\and
APC, 11 Place Marcelin Berthelot, F-75231 Paris Cedex 05, France 
\thanks{UMR 7164 (CNRS, Universit\'e Paris VII, CEA, Observatoire de Paris)}
\and
Dublin Institute for Advanced Studies, 5 Merrion Square, Dublin 2,
Ireland
\and
Landessternwarte, Universit\"at Heidelberg, K\"onigstuhl, D 69117 Heidelberg, Germany
% ...
}

\offprints{J.A.Hinton@leeds.ac.uk, Armand.Fiasson@lpta.in2p3.fr}

\abstract{}{
  The complex Monoceros Loop SNR/Rosette Nebula region contains 
  several potential sources of very-high-energy (VHE) $\gamma$-ray emission
  and two as yet unidentified high-energy EGRET sources.
  Sensitive VHE observations are required to probe acceleration
  processes in this region.
}{
  The H.E.S.S. telescope array has been used to search for
  very high-energy $\gamma$-ray sources in this 
  region. CO data from the NANTEN telescope were used to map the molecular 
  clouds in the region, which could act as target material for 
  $\gamma$-ray production via hadronic interactions.
}{ 
  We announce the discovery of a new $\gamma$-ray source, HESS\,J0632+058,
  located close to the rim of the Monoceros SNR. This source is unresolved 
  by H.E.S.S. and has no clear counterpart at other wavelengths but is possibly
  associated with the weak X-ray source 1RXS\,J063258.3+054857, the Be-star MWC\,148
  and/or the lower energy $\gamma$-ray source 3EG\,J0634+0521.
  No evidence for an associated molecular cloud was found in the CO data.
}{}


\maketitle

\section{Introduction}

Shell-type supernova remnants (SNRs) have been identified as particle
accelerators via their very-high-energy (VHE; $E>100$ GeV)
$\gamma$-ray and non-thermal X-ray emission (see
e.g. \citet{HESS:RXJ1713} and \citet{ASCA:RXJ1713}).  It has been
suggested that interactions of particles accelerated in SNR with
nearby molecular clouds should produce detectable $\gamma$-ray
emission~\cite{ADV1994}. For this reason the well-known Monoceros Loop
SNR (G\,205.5+0.5, distance $\sim$1.6~kpc~\cite{Graham1982,Leahy1986}), 
with its apparent
interaction with the Rosette Nebula (a young stellar cluster/molecular
cloud complex, distance $1.4
\pm 0.1$ kpc~\cite{Hensberge2000}) is a prime target for observations with
VHE $\gamma$-ray instruments. 

For the case of \emph{hadronic} cosmic rays (CRs) interacting in the
interstellar medium to produce pions and hence $\gamma$-rays
via $\pi^{0}$ decay, a spatial correlation between 
$\gamma$-ray emission and tracers of
interstellar gas is expected. Such a correlation was used 
to infer the presence of a population of recently
accelerated CR hadrons in the Galactic Centre region~\cite{HESS:gc_diffuse}.
This discovery highlights the importance of accurate mapping 
of available target material for the interpretation of TeV 
$\gamma$-ray emission. The NANTEN 4~m diameter sub-mm telescope at Las Campanas observatory,
Chile, has been conducting a $^{12}$CO ($J$=1$\rightarrow$0) survey of
the Galactic plane since 1996~\cite{NANTEN:GPS}. The Monoceros region is
covered by this survey and the NANTEN data are used here to trace
the target material for interactions of accelerated hadrons.

\section{H.E.S.S. Observations and Results}

The observations described here took place between March 2004 and
March 2006 and comprise 13.5 hours of data after data quality
selection and dead-time correction. The data were taken over a wide range
of zenith angles from 29 to 59 degrees, leading to a mean energy threshold of 400
GeV with so-called \emph{standard cuts} used here for spectral
analysis and 750~GeV with the \emph{hard cuts} used here for the
source search and position fitting.  These cuts are described in
detail in \citet{HESS:Crab}.

\begin{figure}
  \centering
\hspace{-2.5mm}\resizebox{1.03\hsize}{!}{\includegraphics{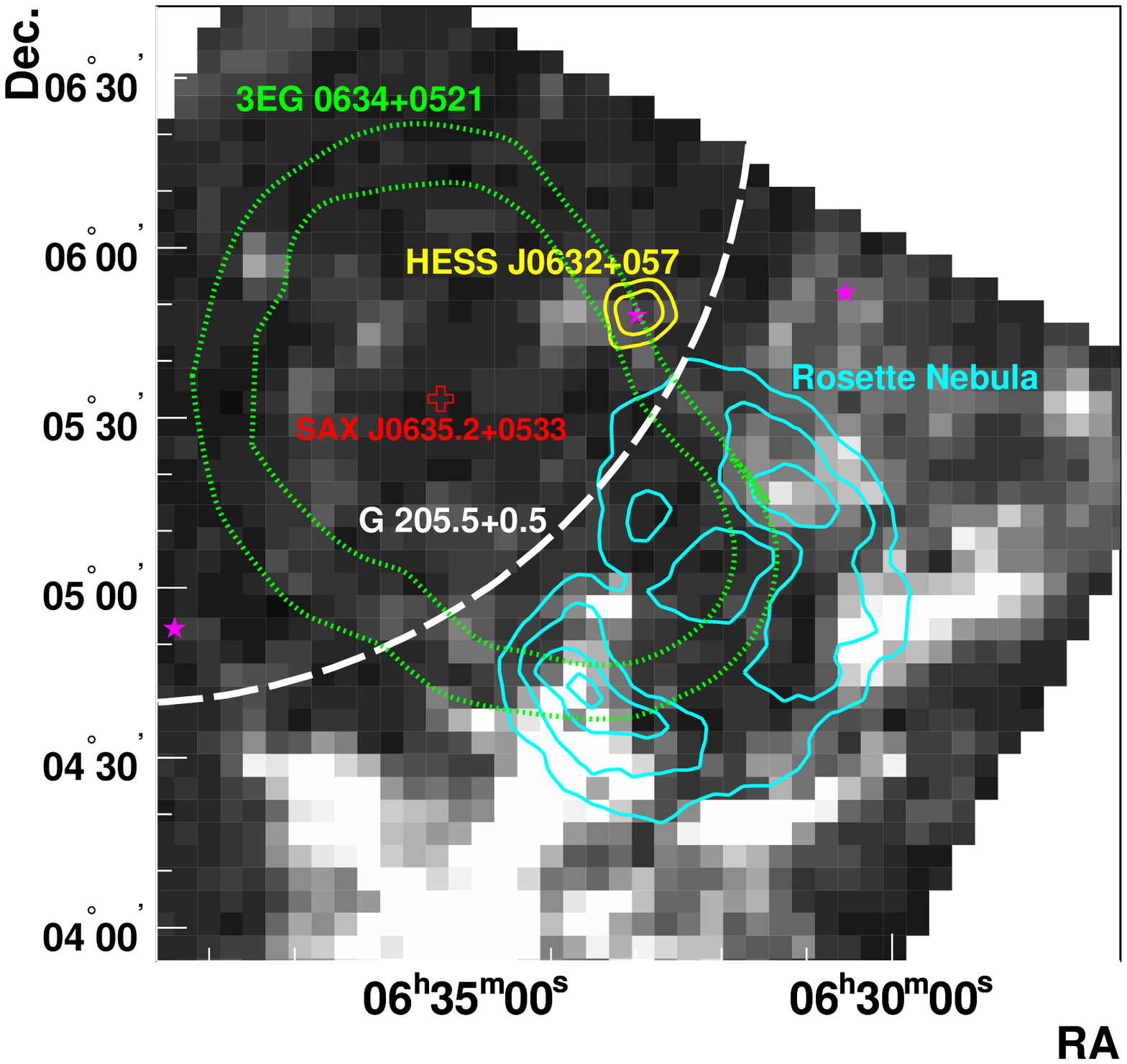}\vspace{-5mm}}
\resizebox{0.99\hsize}{!}{\vspace{-2mm}\includegraphics{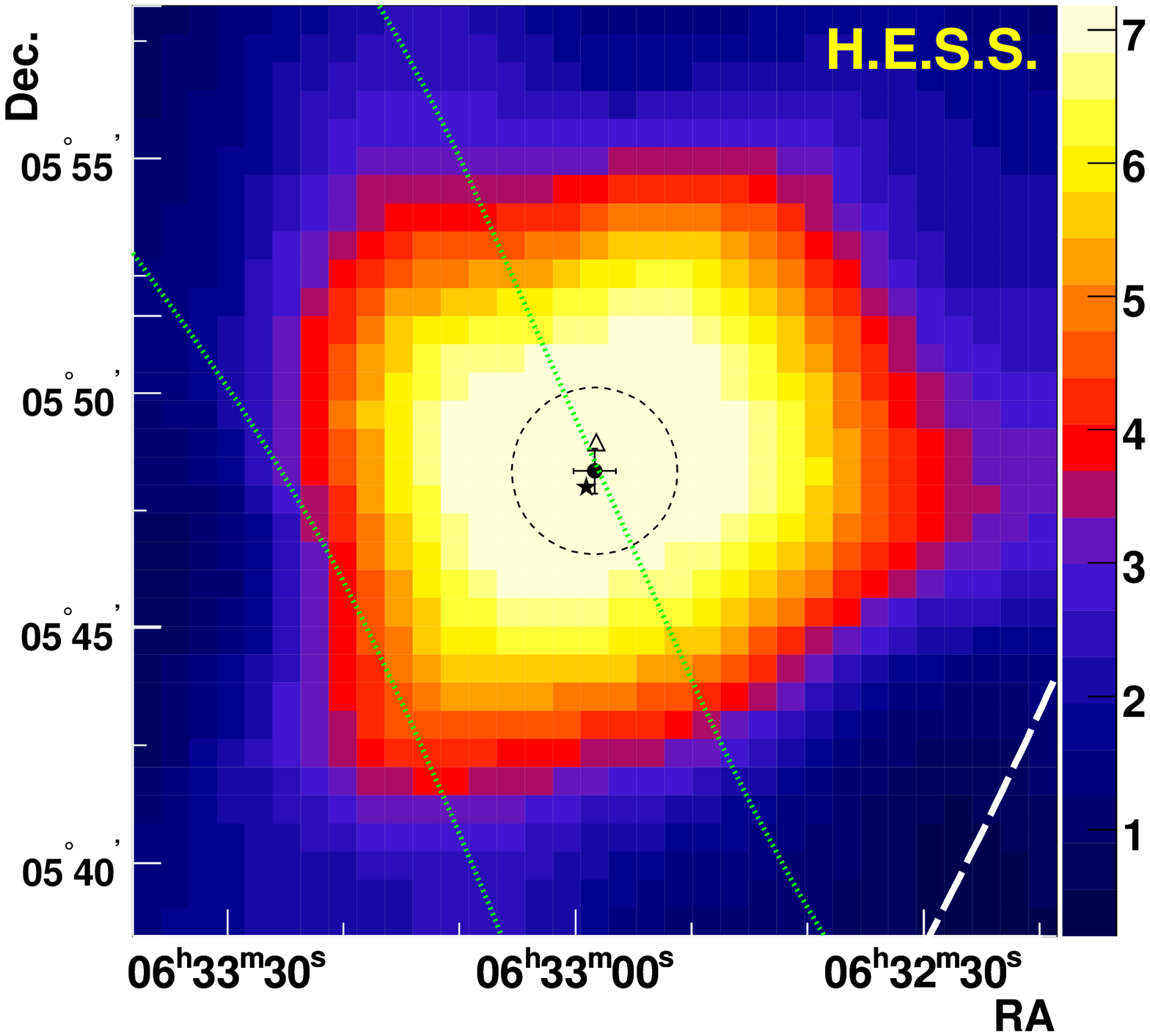}} 
    \caption{ Top: the Monoceros SNR / Rosette Nebula region. 
      The grey-scale shows velocity integrated (0-30 km s$^{-1}$) $^{12}$CO ($J$=1$\rightarrow$0) emission 
      from the NANTEN Galactic Plane Survey (white areas have highest flux). 
      Yellow contours show
      4 and 6 $\sigma$ levels for the statistical significance of a point-like
      $\gamma$-ray excess. 
      Radio observations at 8.35~GHz from \citet{Radio:GPS8GHz} are
      overlaid as cyan contours, and illustrate the
      extent of the Rosette Nebula. The nominal \citet{Green2004} Catalogue
      position/size of the Monoceros SNR is shown as an (incomplete)
      dashed circle. 95\% and 99\% confidence regions for the position of
      the EGRET source 3EG\,0634+0521 are shown as dotted green contours.
      The binary pulsar SAX\,J0635.2+0533 is marked with a square and 
      Be-stars with pink stars.
      Bottom: an expanded view of the 
      centre of the top panel showing H.E.S.S. significance as
      a colour scale. The rms size limit derived for the TeV 
      emission is shown as a dashed circle. The unidentified X-ray source 
      1RXS\,J063258.3+054857 is marked with a triangle and the Be-star
      MCW~148 with a star.
    }
    \label{fig:skymap}
\end{figure}

A search in this region for point-like emission was made using a
0.11$^{\circ}$ \emph{On source} region and a ring of mean radius 
0.5$^{\circ}$ for \emph{Off source} background estimation (see \citet{Berge:BG} for
details). Fig.~\ref{fig:skymap}
shows the resulting significance map, together with CO data from NANTEN,
radio contours and the positions of all Be-stars in this region. The peak significance in the
field is $7.1\sigma$. 
The number of statistical trials associated with a search of
the entire field of view, in $0.01^{\circ}$ steps along both axes, is
$\approx 10^{5}$. The measured peak significance corresponds 
to $5.3\sigma$ after accounting for these trials. 
A completely independent analysis based on
a fit of camera images to a shower model (\emph{Model Analysis} described in
\citet{deNaurois:Model}), yields a
significance of $7.3\sigma$ ($5.6\sigma$ post-trials).  

\begin{figure}
  \vspace{-2mm}
  \centering \resizebox{0.98\hsize}{!}{\includegraphics{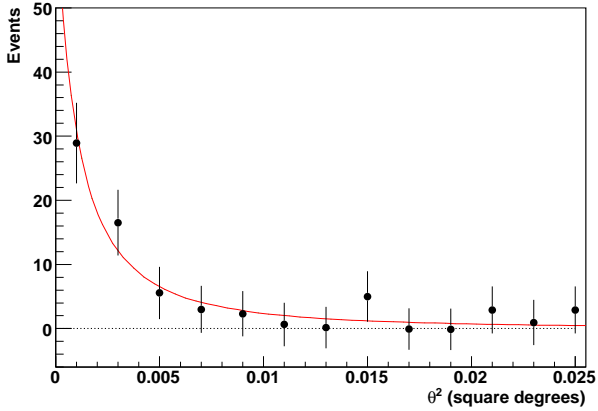}} 
\caption{
   Distribution of excess (candidate $\gamma$-ray) events as a function of squared angular
   distance from the best fit position of HESS\,J0632+057 (points),
   compared to the expectation for this dataset from Monte-Carlo
   simulations (smooth curve). 
  }
  \label{fig:thsq}
\end{figure}

The best fit position of the new source is 
$6^{\mathrm{h}}32^{\mathrm{m}}58.3^{\mathrm{s}}$, $+5^{\circ}48'20''$ 
(RA/Dec. J2000) with 28$''$ statistical
errors on each axis, and is hence identified as HESS\,J0632+057. 
Systematic errors are estimated at 20$''$ on each
axis.  There is no evidence for intrinsic extension of the source and
we derive a limit on the rms size of the emission region of $2'$ (at
95\% confidence), under the assumption that the source follows a
Gaussian profile.  This source size upper limit is shown as a dashed
circle in the bottom panel of Fig.~\ref{fig:skymap}.
Fig.~\ref{fig:thsq} demonstrates the point-like nature of the
source.  The angular distribution of excess $\gamma$-ray-like events
with respect to the best fit position is shown together with the
expected distribution for a point-like source.

\begin{figure}[h]
  \vspace{-2mm}
  \centering \resizebox{0.98\hsize}{!}{\includegraphics{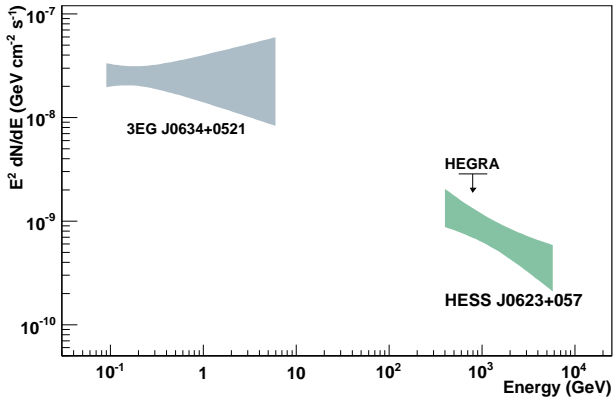}}
  \caption{ Reconstructed VHE $\gamma$-ray spectrum of HESS\,J0632+057
    compared to the HE $\gamma$-ray source 3EG\,J0634+0521. An upper
    limit derived for 3EG\,J0634+0521 at TeV energies using the HEGRA
    instrument is also shown.
}
  \label{fig:spectrum}
\end{figure}

The reconstructed energy spectrum of the source is consistent with a
power-law: $dN/dE = k (E/ 1 \mathrm{TeV})^{-\Gamma}$ with photon index
$\Gamma = 2.53\pm0.26_{stat}\pm0.20_{sys}$ and a flux normalisation $k
= 9.1\pm1.7_{stat}\pm3.0_{sys} \times 10^{-13}$ cm$^{-2}$s$^{-1}$TeV$^{-1}$.  
Fig.~\ref{fig:spectrum} shows the H.E.S.S.  spectrum together
with that for the unidentified EGRET source 3EG\,J0634+0521 (discussed
below) and an upper limit derived for TeV emission from 3EG\,J0634+0521
using the HEGRA telescope array~\cite{HEGRA:monoceros}, converted from
an integral to a differential flux using the 
spectral shape measured by H.E.S.S. 
We find no evidence for flux variability of HESS\,J0632+057 within our dataset.
However, we note that due to the weakness of the source and sparse sampling of the light-curve,
intrinsic variability of the source is not strongly constrained.
The bulk of the available data was taken in two short periods in
December 2004 (P1, 4.7 hours) and November/December 2005 (P2, 6.2 hours).
The integral fluxes (above 1 TeV) in these two periods were: 
$6.3 \pm 1.8 \times 10^{-13}$ cm$^{-2}$ s$^{-1}$ (P1) and 
$6.4 \pm 1.5 \times 10^{-13}$ cm$^{-2}$ s$^{-1}$ (P2).

Amongst the candidate VHE sources in this field is the 34~ms binary
pulsar SAX\,J0635.2+0533.  There is no significant $\gamma$-ray
emission at the position of this object and we derive a 99\% confidence
upper limit on the integral flux, $F(>1 \mathrm{TeV})$, of $2.6 \times 10^{-13}$ cm$^{-2}$
s$^{-1}$, assuming an $E^{-2}$ type spectrum.

\section{Possible Associations of HESS\,J0632+057}

The new VHE source HESS\,J0632+057 lies in a complex region and
several associations with objects known at other wavelengths seem
plausible. We therefore consider each of these potential 
counterparts in turn.

{\bf The Monoceros Loop SNR} is rather old in comparison to the 
known VHE $\gamma$-ray shell-type SNRs RX\,J1713.7$-$3946
\cite{HESS:RXJ1713}, RX\,J0852.0$-$4622~\cite{HESS:velajnr}
and Cas-A~\cite{HEGRA:CasA}. All these objects have estimated
ages less than $\sim2000$ years, in contrast the Monoceros
Loop SNR has an age of $\sim3\times 10^{4}$ years
\cite{Leahy1986}.
This supernova remnant therefore appears to be in a different
evolutionary phase (late Sedov or Radiative) compared to these known
VHE sources. However, CR acceleration may occur 
even at this later evolutionary stage (see for example \citet{Yamazaki06}). 
The principal challenge for a scenario involving the Monoceros Loop is to 
explain the very localised VHE emission at only one point
on the SNR limb. The interaction of the SNR with a compact
molecular cloud is one possible solution.
In this scenario (and indeed any $\pi^{0}$ decay scenario) for the observed $\gamma$-ray emission, a 
correlation is expected between the TeV emission and the distribution of
target material. An unresolved molecular cloud listed in a CO survey at
115~GHz~\citep{Oliver1996} lies rather close to HESS\,J0623+057, at
$l=205.75$ $b=-1.31$. The distance estimate for this cloud (1.6 kpc)
is consistent with that for the Monoceros SNR, making it a potential
target for hadrons accelerated in the SNR. However, as can be seen
clearly in the NANTEN data in Fig.~\ref{fig:skymap}, 
the intensity peak of this cloud is
significantly shifted to the East of the H.E.S.S. source. 
We find no evidence in the NANTEN data for any clouds along the line of
sight to the H.E.S.S. source.

{\bf 3EG\,J0634+0521} is an unidentified EGRET source \cite{EGRET:3EG}
with positional uncertainties such that HESS\,J0632+057 lies 
close to the 99\% confidence contour. 
Given that this source is flagged as possibly extended or confused, 
a positional coincidence of these two objects seems plausible.
Furthermore, the reported third EGRET catalogue flux above 100
MeV ($(25.5\pm5.1)\times10^{-8}$ photons cm$^{-2}$ s$^{-1}$ with a
photon index of $2.03\pm0.26$, see Fig.~\ref{fig:spectrum}), is
consistent with an extrapolation of the H.E.S.S. spectrum. A global
fit of the two spectra gives a photon index of 2.41$\pm$0.06. 

{\bf 1RXS\,J063258.3+054857} is a faint ROSAT source
\cite{ROSATFaint} which lies 36$''$ from the H.E.S.S. source 
with a positional uncertainty of 21$''$
(see Fig.~\ref{fig:skymap} bottom).
Given the uncertainties on the positions of both objects this X-ray
source can certainly be considered a potential counterpart of 
HESS\,J0632+057. The chance probability of the coincidence of a
ROSAT Faint Source Catalogue source within the H.E.S.S. error circle 
is estimated as 0.1\% by scaling the total number of sources in 
the field of view. The ROSAT source is rather weak, with 
only 4 counts detected above 0.9~keV, spectral comparison is 
therefore rather difficult. In the scenario where the $\gamma$-ray
emission is interpreted as inverse Compton emission from a 
population of energetic electrons, the ROSAT source could be naturally
ascribed to the synchrotron emission of the same electron population.
However, the low level of the X-ray emission ($\sim10^{-13}$ erg cm$^{-2}$ s$^{-1}$)
in comparison with the TeV flux ($\sim10^{-12}$ erg cm$^{-2}$ s$^{-1}$)
implies a very low magnetic field ($\ll 3 \mu$G) unless a strong radiation  
source exists in the neighbourhood of the emission region and/or
the X-ray emission suffers from substantial absorption. Observations at 
$>4$ keV are required to resolve this absorption issue.
In a $\pi^{0}$ decay scenario for the $\gamma$-ray source, 
secondary electron production via muon decay
is expected along with $\gamma$-ray emission. The synchrotron emission 
of these secondary electrons would in general produce a weaker X-ray source
than the IC scenario, probably compatible with the measured ROSAT flux.

{\bf MWC\,148} (HD\,259440) is a massive emission-line star of spectral
type B0pe which lies within the H.E.S.S. error circle. The chance
probability of this coincidence is hard to assess, as there was no
a-priori selection of stellar objects as potential $\gamma$-ray sources.
However, given the presence of only 3 Be-type stars in the field of view of the
H.E.S.S. observation (see Fig.~\ref{fig:skymap}) and the solid angle
of the H.E.S.S. error circle, the naive chance probability of the
association is $10^{-4}$. 
Stars of this spectral type have winds with typical
velocities and mass loss rates of 1000~km s$^{-1}$ and $10^{-7}
M_{\odot}/$year, respectively.
Plausible acceleration sites are in strong internal or external shocks
of the stellar wind. We estimate that an efficiency of 1-10\% in the
conversion of the kinetic energy of the wind into $\gamma$-ray
emission would be required to explain the H.E.S.S. flux (assuming this
star lies at the distance of the Rosette Nebula).
However, as no associations of similar stars with point-like $\gamma$-ray
sources were found in the H.E.S.S. survey of the inner Galaxy,
this scenario seems rather unlikely.

A related possibility is that MWC\,148 is part of a binary system
with an, as yet undetected, compact companion. Such a system might
then resemble the known VHE $\gamma$-ray source PSR\,B1259-63/SS\,2883~\cite{HESS:psrb1259}.
Further multi-wavelength observations are required to confirm or
refute this scenario.

\begin{acknowledgements}

The support of the Namibian authorities and of the University of
Namibia in facilitating the construction and operation of H.E.S.S. is
gratefully acknowledged, as is the support by the German Ministry for
Education and Research (BMBF), the Max Planck Society, the French
Ministry for Research, the CNRS-IN2P3 and the Astroparticle
Interdisciplinary Programme of the CNRS, the U.K. Particle Physics and
Astronomy Research Council (PPARC), the IPNP of the Charles
University, the South African Department of Science and Technology and
National Research Foundation, and by the University of Namibia. We
appreciate the excellent work of the technical support staff in
Berlin, Durham, Hamburg, Heidelberg, Palaiseau, Paris, Saclay, and in
Namibia in the construction and operation of the equipment. The
NANTEN project is financially supported from JSPS (Japan Society for
the Promotion of Science) Core-to-Core Program, MEXT Grant-in-Aid for
Scientific Research on Priority Areas, and SORST-JST (Solution
Oriented Research for Science and Technology: Japan Science and
Technology Agency). We would also like to thank Stan Owocki and James
Urquhart for very useful discussions.

\end{acknowledgements}

\bibliographystyle{aa} 
\bibliography{mono}

\end{document}